\documentclass[fleqn,10pt]{wlscirep}
\usepackage[utf8]{inputenc}
\usepackage[T1]{fontenc}
\usepackage{float}
\title{Volumetric Non-Invasive Cardiac Mapping for Accessible Global Arrhythmia Characterization}
\author[1,2,*]{Jorge Vicente-Puig}
\author[1]{Judit Chamorro-Servent}
\author[2]{Ernesto Zacur}
\author[3]{Inés Llorente-Lipe}
\author[3]{Marta Martínez}
\author[4,5,6]{Jorge Sanchez}
\author[2,3,7]{Jana Reventós}
\author[7]{Ivo Roca-Luque}
\author[7]{Lluis Mont}
\author[8,9]{Felipe Atienza}
\author[2,3]{Andreu M. Climent}
\author[2,3]{Maria S. Guillem}
\author[2,3]{Ismael Hernández-Romero}

\affil[1]{Universitat Autònoma de Barcelona, Spain}
\affil[2]{Corify Care S.L, Madrid, Spain}
\affil[3]{ITACA Institute, Universitat Politècnica de València, Valencia, Spain}
\affil[4]{Centro de Investigación e Innovación en Bioingeniería, Universidad Politecnica de Valencia, Valencia, Spain}
\affil[5]{Institute of biomedical engineering, Karlsruhe Institute of Technology, Karlsruhe, Germany}
\affil[6]{Universidad Internacional de Valencia, Valencia, Spain}
\affil[7]{Arrhythmia Section, Cardiology Department, Hospital Clínic, Universitat de Barcelona, Barcelona, Catalonia, Spain}
\affil[8]{Department of Cardiology, Hospital General Universitario Gregorio Marañón, Instituto de Investigación Sanitaria Gregorio Marañón (IISGM), Madrid, Spain}
\affil[9]{Centro de Investigación Biomédica en Red de Enfermedades Cardiovasculares (CIBERCV), Madrid, Spain}
\affil[*]{Corresponding author, jorge.vicente@corifycare.com}

\usepackage{lineno}
\usepackage[normalem]{ulem}
\usepackage{etoolbox}                

\newenvironment{plainlanguagesummary}[1][Plain Language Summary]{%
  \par\medskip
  \noindent\textbf{#1}\par
  \begingroup\small\setlength{\parindent}{0pt}%
}{%
  \par\endgroup\medskip
}

\usepackage{comment}

\usepackage{graphicx}

\begin{abstract}
\textbf{Introduction:} Cardiac arrhythmias are a major cause of morbidity and mortality increasing the risk of stroke, heart failure, and sudden cardiac death. Imageless electrocardiographic Imaging has emerged as an accessible non-invasive alternative for cardiac electrical mapping from body surface potentials. However, conventional electrocardiographic imaging is restricted to epicardial reconstructions, reducing its reliability in accurately identifying arrhythmias arising from deeper myocardial structures. We aim to overcome this limitation by reconstructing three-dimensional cardiac activity. \\\\
\textbf{Methods:} We introduce a volumetric formulation, which extends beyond epicardial potential estimation by solving an inverse source problem using Green’s functions. This technique enables three-dimensional reconstructions of cardiac activation, improving arrhythmia localization in anatomically complex regions. We evaluate the method on simulated premature ventricular beats and on four patients representing clinical challenges, including a premature ventricular contraction from the right ventricular outflow tract, a left bundle branch block, a ventricular tachycardia, and a Wolff-Parkinson-White. We also assess performance on an open-source dataset for myocardial infarction estimation. \\\\
\textbf{Results:} Our results indicate that volumetric electrocardiographic imaging reconstructs three-dimensional activation and enhances the localization of arrhythmia origins, yielding a 59.3\% reduction in geodesic error between the estimated and simulated origins compared to surface-only approaches. In patient cases, the recovered activation patterns are consistent with the clinical diagnoses. \\\\
\textbf{Conclusions:} Imageless volumetric electrocardiographic imaging enables non-invasive, accessible, three-dimensional mapping of cardiac activation, addressing a fundamental limitation of surface-restricted methods. This capability may support more accurate pre-procedural planning, may help guide ablation targets, and could refine selection and optimization of cardiac resynchronization therapy candidates.
\end{abstract}

\begin{document}

\flushbottom
\maketitle
\thispagestyle{empty}

\begin{plainlanguagesummary}
Heart rhythm disorders are common and often require invasive procedures to diagnose and treat. To reduce that need, advanced non-invasive methods such as electrocardiographic imaging (ECGI) use signals from chest electrodes, the person’s body geometry, and a physics-based computational model to create maps of the heart’s electrical activity. Most current ECGI maps cover only the heart’s surface. We developed a non-invasive approach that maps activation within the heart muscle in three dimensions. We tested it in computer simulations, in four patients with representative rhythm problems, and on a public dataset related to prior cardiac infarctions. We show that it identifies where abnormal beats start more accurately than surface-only maps and agrees with clinical assessments. In future studies with larger cohorts will aim to evaluate whether this technique can improve procedure planning, guide ablation targets, and aid in selecting candidates for resynchronization therapy.
\end{plainlanguagesummary}

\section*{Introduction}
Cardiac arrhythmias affect one in three adults\cite{hindricks20212020} and pose significant health risks, including stroke, heart failure, and sudden cardiac death. Conditions such as atrial fibrillation (AF), ventricular tachycardia (VT), and conduction disorders often require medical or interventional treatments. These conditions contribute substantially to healthcare costs, accounting for 2–4\% of total annual healthcare expenditure in Europe\cite{HFPN2020} and resulting in an annual cost exceeding €13.5 billion\cite{kirchhof20162016, sanchez2023cost}. These challenges, which are particularly acute in resource-constrained environments facing limited access to specialized diagnostics and care, emphasize the need for improved diagnosis and treatment strategies.\\\\
Current treatments, including anti-arrhythmic medications and catheter-based ablation, have limited efficacy and notable drawbacks. Medications often fail due to low success rates and side effects\cite{calkins2009treatment}, while electroanatomical mapping and ablation require invasive procedures with risks, repeat interventions, and limited insight into epicardial or mid-myocardial activity\cite{oral2006circumferential}. These limitations underscore the urgent need for advanced, non-invasive mapping technologies to enhance arrhythmia diagnosis, guide treatment, and improve clinical outcomes. \\

In recent years, Electrocardiographic Imaging (ECGI) has emerged as a promising non-invasive technology for estimating the electrical activity of the heart by recording body surface potentials from multiple electrodes placed on the torso\cite{hernandez2023electrocardiographic, cluitmans2018validation}.
Unlike standard electrocardiogram (ECG), which provides only indirect information about cardiac activation sequences, ECGI allows estimating the distribution of electrical potentials over the epicardial surface, offering a high-resolution map of arrhythmic activity.
This technique has been particularly valuable for localizing atrial and ventricular arrhythmogenic substrates, aiding in the identification of focal and reentrant circuits in conditions such as AF\cite{knecht2017multicentre}, atrial tachycardia (AT)\cite{shah2013validation}, VT\cite{wang2011noninvasive}, and premature ventricular contractions (PVCs)\cite{yu2017three} among others\cite{pereira2020electrocardiographic}. Additionally, ECGI has played a role in guiding personalized ablation strategies and in assessing conduction disorders such as Left Bundle Branch Block (LBBB) and ventricular dyssynchrony\cite{pereira2020electrocardiographic}. The clinical relevance of ECGI has recently been acknowledged in clinical guidelines, particularly for its potential role in the management of AF\cite{tzeis20242024} and CRT\cite{euaf050}.\\\\
One of the main limitations of ECGI systems is their reliance on cardiac imaging modalities, such as computed tomography (CT) or magnetic resonance imaging (MRI), to accurately map electrode positions onto cardiac geometries. This requirement not only increases procedural time but also adds significant costs, posing a major barrier to clinical adoption, especially in resource-constrained environments. To mitigate the need for CT/MRI ECGI systems capable of estimating cardiac geometry without dedicated imaging, known as Imageless ECGI, have been developed \cite{molero2023robustness, jove67958}. These systems facilitate the rapid generation of cardiac maps, improving accessibility and cost-effectiveness. Our volumetric ECGI pipeline builds on these imageless systems and therefore enables volumetric cardiac estimation without the additional costs of cardiac imaging, while running on standard hardware. These features facilitate adoption in diverse clinical contexts, with particular relevance to resource-constrained settings.\\

However, both traditional and imageless ECGI typically remain limited to epicardial reconstructions\cite{cheniti2019noninvasive, reventos2022validation}. This restricts their clinical utility when arrhythmogenic substrates reside deeper within the myocardium, making it difficult to accurately resolve activation details crucial for certain conditions. For instance, while ECGI aids in localizing PVCs\cite{yu2017three}, precisely identifying their origin from complex, deep structures like the septal region of the Right Ventricular Outflow Tract (RVOT) remains challenging using only surface-based epicardial potentials\cite{cluitmans2018validation, campos2024silico}. Similarly, assessing LBBB\cite{pereira2020electrocardiographic} with epicardial ECGI provides limited insight into the full transmural activation sequence relevant for cardiac resynchronization therapy optimization.
Mapping of ventricular tachycardia (VT) circuits with ECGI has been explored, particularly to guide noninvasive ablation procedures~\cite{sapp2020mapping,graham2021use}, yet these studies have focused almost exclusively on epicardial reentrant pathways, offering little insight into deeper intramural circuits.
Furthermore, identifying accessory pathways (e.g., in Wolff-Parkinson-White (WPW) syndrome) often relies predominantly on ECG interpretation or invasive studies, as epicardial mapping alone may not capture the complete conduction pattern\cite{pereira2020electrocardiographic}. Although strategies extending ECGI beyond the epicardium, using action potentials\cite{messnarz2004new}, electric potentials\cite{kalinin2019solving}, or other inverse methods\cite{ondrusova2023two, wang2013inverse} have been proposed, they lack widespread clinical adoption. Reliably reconstructing activation throughout the myocardial volume remains a significant technical hurdle limiting their application. \\

To overcome these limitations, we propose a novel volumetric ECGI approach, by explicitly modelling cardiac sources within the Poisson-based formulation, enabling the estimation of three-dimensional activation patterns throughout the entire myocardium. This inherently allows for the simultaneous reconstruction of activation across multiple cardiac chambers (e.g., atria and ventricles). This methodology is based on Green’s functions to describe the transfer between body surface potentials and myocardial electrical activity, providing a more comprehensive understanding of cardiac excitation patterns in a complete representation of the heart. \\

In this study, we present a volumetric electrocardiographic imaging (ECGI) approach to estimate cardiac sources, enabling three-dimensional reconstruction of activation throughout the myocardium. The method is validated on simulations of premature ventricular contractions (PVCs) initiated at multiple sites, and it localizes the origin of abnormal beats more accurately than a classical surface-based ECGI approach. On patient data of different pathologies and on an open myocardial infarction dataset, it recovers activation patterns and infarct-related abnormalities that align with clinical assessments. These results indicate that volumetric ECGI provides reliable non-invasive mapping across diverse scenarios. This capability may support more precise pre-procedural planning, guides ablation targets, and may refine selection for cardiac resynchronization therapy, while potentially reducing reliance on invasive procedures, an especially relevant benefit in resource-limited settings.

\section*{Methods}
\begin{figure}[h]
    \centering
    \includegraphics[width=\linewidth]{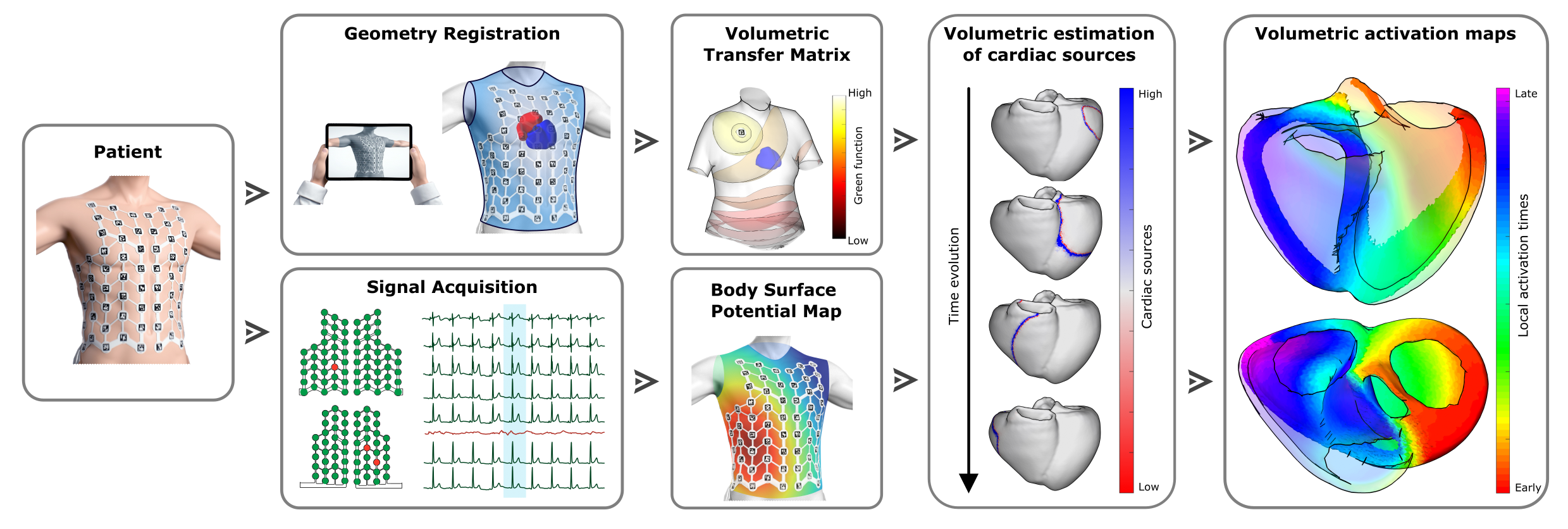}
    \caption{Workflow of volumetric ECGI technology. The process begins with body surface potential acquisition using a multi-electrode vest, capturing electrical signals from the torso. Simultaneously, patient-specific cardiac and torso geometries are estimated. These data are processed through a volumetric transfer matrix, enabling the estimation of cardiac sources throughout the myocardium. Finally, local activation times are computed, generating three-dimensional activation maps that depict the electrical wave propagation within the heart.}
    \label{fig:pipelineVolumetric}
\end{figure}

The volumetric ECGI system allows for the reconstruction of three-dimensional cardiac electrical activity by integrating body surface potential recordings with patient-specific anatomical models. This process consists of  geometry registration, signal acquisition, computation of the volumetric transfer matrix, and estimation of cardiac sources, ultimately leading to the generation of activation maps (Figure \ref{fig:pipelineVolumetric}). \\

To acquire the input data, including both electrical signals and geometries, we use the ECGI system (ACORYS MAPPING SYSTEM, Corify Care SL, Madrid, Spain). A 128-electrode vest is placed on the patient's torso, ensuring stable and accurate measurements. Each electrode allows the recording surface potentials providing a high spatial resolution at the torso, facilitating the reconstruction of cardiac activity. However, some electrodes may be excluded due to excessive noise or poor contact, based on clinical considerations. To enhance signal quality, noise filtering and baseline correction are applied, improving the signal-to-noise ratio (SNR) and ensuring robust data for further processing.\\\\
Once the signals are recorded, the next step is to define the anatomical structures required for cardiac activity estimation. Both the external torso surfaces and internal cardiac structures, such as the atria and ventricles, play a crucial role in this process. The torso geometry is reconstructed using photogrammetry, capturing electrode positions and surface topology, which serves as the foundation for subsequent cardiac modelling. Then, cardiac geometry is estimated through a Statistical Shape Model (SSM)\cite{zacur2017mri}, generating a patient-specific representation of the heart. While simulations inherently provide detailed internal geometries, patient-specific reconstructions rely on shape models to estimate cardiac structures in real-time. By removing the reliance on CT or MRI scans, this approach expands its clinical applicability. \\\\
Upon defining the geometry, a transfer matrix is computed to describe the relationship between surface potentials and myocardial electrical activity\cite{malmivuo1995bioelectromagnetism, barr1977relating}, which allows for the cardiac source reconstruction. Traditional ECGI methodologies, often referred to as epicardial ECGI, rely on the theory of electrical passive conductors and thus solve a Cauchy problem for Laplace’s equation to estimate potentials on intermediate surfaces between the heart and torso\cite{barr1977relating}. 
In contrast, the proposed volumetric ECGI is formulated as a linear inverse source problem, explicitly estimating cardiac sources that represent the propagating wavefront through the entire heart. This methodology is grounded in the theory of electric volume conductors, described by Poisson’s equation, and relies on Green’s functions\cite{franklin2012green} to model the transfer between electrode potentials and cardiac sources. Consequently, this approach enables a three-dimensional estimation of electrical activity across epicardial, endocardial, and myocardial regions. Moreover, this formulation allows the incorporation of anatomically detailed chambers, such as both atria and ventricles, ensuring a more comprehensive representation of cardiac structures. \\\\
To further refine the reconstruction, an optimization algorithm, such as Tikhonov regularization\cite{hansen2010discrete}, is applied to identify the most plausible electrical activity that matches the electrode recordings. This process allows to compute clinically relevant information, including Local Activation Times (LATs) in regular rhythms. The mathematical formulation and implementation of this inverse problem are detailed in the following subsections.

\subsection*{Computational dataset}
This study was validated using both simulated and patient data, incorporating a dataset of 16 simulated Premature Ventricular Contractions (PVCs) and four patients diagnosed with PVCs, Left Bundle Branch Block (LBBB), Ventricular Tachycardia (VT) and Wolff-Parkinson-White (WPW) syndrome . \\\\
Simulated data was generated using realistic anatomical ventricular meshes aligned with torso geometries according to anatomical constraints\cite{qian2023additional, sanchez2025enhancing}, incorporating structures like lungs, blood pools, and the liver. The extracellular medium was considered isotropic, and conductivity for the blood, lungs, liver, and torso were $0.7 S/m$, $0.0389 S/m$, $0.1667 S/m$, and $0.8 S/m$, respectively\cite{qian2023additional}. Ventricular mesh, with an average edge length of $350 \ \mu \text{m}$, included myocyte orientation modeled via rule-based methods\cite{bayer2012novel}. The bidomain model was solved using openCARP\cite{plank13opencarp} to simulate the electrical propagation in the ventricular myocardium and the electrical fields across the torso. \\\\
For ECGI computation, a reduced torso surface with an average edge length of $13.22 \pm 2.41$ mm was used. epicardial ECGI utilized a reduced ventricular surface ($2.28 \pm 0.38$ mm edge length), whereas volumetric ECGI employed a tetrahedral ventricular myocardial mesh ($3.53 \pm 0.75$ mm edge length). 128 electrode locations were selected to match the vest placement, leading to a dataset of 16 PVC simulations: 6 originating from the ventricular base, 6 from the septum, and 4 from the free wall. 

\subsection*{Clinical dataset}
Four patients (ages 58–80) were registered with ECGI using a 128-lead system (ACORYS MAPPING SYSTEM, Corify Care SL, Madrid, Spain). Patient 1 was registered at Hospital Gregorio Marañón and patients 2, 3 and 4 were registered at Hospital Clínic de Barcelona . The study received ethical approval, and informed consent was obtained. All patients were registered under the study protocol SAVE-COR (NCT05772182).\\\\
Patient 1, a 62-year-old individual, presented with ventricular extrasystole, which was initially managed with beta-blockers. The arrhythmia was successfully eliminated following catheter ablation in the right ventricular outflow tract (RVOT). Patient 2, an 80-year-old individual, was diagnosed with Left Bundle Branch Block (LBBB) and underwent cardiac resynchronization therapy (CRT). ECGI mapping captured the conduction pattern, and monopolar and biventricular stimulation were performed to achieve ventricular synchronization. Patient 3, a 51-year-old individual, presented with sustained ventricular tachycardia. Preprocedural fibrosis substrate detection using ADAS 3D identified a fibrotic region located on the basal part of the septal wall of the left ventricle. Patient 4, a 58-year-old individual, was diagnosed with Wolff-Parkinson-White (WPW) syndrome, characterized by an accessory pathway bypassing the Purkinje system. The pathway was localized through catheter analysis in the postero-septal region of the ventricle, confirming the diagnosis. \\\\
To enhance comparison with previous studies and explore the viability of myocardial infarction detection two patients from the 2007 Physionet Challenge of Computation of Cardiology congress were included\cite{goldberger2000physiobank}. Body surface potential recordings were obtained at 120 anatomical sites, following the protocol established by Dalhousie University. Each dataset consisted of a single averaged PQRST complex, sampled at 2 kHz. Ventricular and torso geometries were extracted and processed using proprietary software.

\subsection*{ECGI system}
Although simulations provide exact geometries and electrical signals, patient data require a detailed description of the ECGI system used for acquisition. ECGI relies on two primary types of input data: geometrical information and electrical signals. Geometrical information consists of patient-specific torso and heart surfaces, enabling accurate reconstruction of cardiac activity. Electrical signals are recorded from electrodes placed on the torso, providing a high-resolution mapping of surface electrical activity. \\\\
The sensor vest is a high-density electrode array with 128 silver electrodes that enable simultaneous mapping of surface electrical activity across the patient’s entire torso. The vest is radiolucent and contains four patches that cover the anterior and posterior left and right surfaces of the torso. Each electrode features a QR code on the front, allowing for automatic electrode position identification. The vest is flexible, allowing custom adaptation to individual torso shapes and modifications for compatibility with other electrophysiological systems used in clinical practice. Once the vest is in place, photogrammetry scanning is performed to obtain a detailed 3D representation of the patient's torso, including electrode positions which are automatically located with the QR code. This torso representation is then converted into a triangular mesh, forming the basis for subsequent numerical computations. The cardiac geometry, estimated using a Statistical Shape Model (SSM)\cite{zacur2017mri}, is then converted into one or multiple triangular meshes for further computational processing.\\\\
The electrode vest is connected to the biopotential amplifier, which is an isolated 128-channel device responsible for amplifying and digitizing the electrical signals collected by the electrodes. These signals form a Body Surface Potential Map (BSPM), providing high spatial resolution of electrical activity on the torso surface. Electrodes with excessive noise or poor contact may be excluded based on clinical considerations. Then, the recorded signals are first processed with a comb filter at 50 Hz, which eliminates powerline interference and its three first harmonics. Next, a 10th-order Butterworth low-pass filter at 40 Hz is applied to attenuate higher-frequency noise. Finally, a high-pass filter at 0.67 Hz is used to remove baseline wander. For simulated data, Gaussian noise is added at a 20 dB signal-to-noise ratio (SNR), followed by 10th-order Butterworth low-pass filter with a cut-off frequency of 50Hz. 

\subsection*{Epicardial reconstruction}
Electrocardiographic Imaging (ECGI) is formulated as an inverse problem in which body surface potentials are used to estimate cardiac electrical activity. Conventional ECGI approaches primarily rely on surface-based estimations of the electric potential. Due to the inherent challenges in accurately estimating the electric potential beyond the epicardium, we focus on reconstructing epicardial potentials.\\\\ 
The body is modelled as a homogeneous, quasi-static volume conductor with uniform conductivity to simplify the mathematical formulation. Since the heart is excluded from the domain of interest, there are no primary electrical sources within this region. Under these assumptions, the inverse problem of reconstructing epicardial potentials from body surface potentials is formulated as a Cauchy problem for Laplace’s equation:
\begin{align}  
-\nabla^2 \phi(x) &= 0, \quad x \in \Omega_B, \label{eq:CauchyProblem_elliptic}\\ 
\phi(x) &= g(x), \quad x \in \Gamma, \label{eq:CauchyProblem_Dirichlet}\\ 
\partial_{n_T} \phi(x) &= 0, \quad x \in \partial \Omega, \label{eq:CauchyProblem_Neumann} 
\end{align}
where $\phi$ denotes the electric potential, $g$ represents the recorded or simulated body surface potentials, and $\Omega_B$ is the domain enclosed between the torso surface $T$ and the heart surface $H$, such that $\partial \Omega_B = T \cup H$ and $T \cap H = \emptyset$. The set $\Gamma \subset \partial \Omega$ corresponds to the electrode locations, $\partial \Omega = T$ to the torso surface, and $\partial_{n_T}$ represents the normal derivative at the torso surface.  \\\\
To solve this problem, we discretize the Cauchy problem using the Boundary Element Method (BEM)\cite{barr1977relating}, yielding a transfer matrix that relates the torso potentials to heart potentials via the linear system given:
\begin{equation}
    A h = g,
\end{equation}
where $A \in \mathbb{R}^{M \times N}$ is the transfer matrix, $h \in \mathbb{R}^{N \times T}$ represents the heart potentials, and $g \in \mathbb{R}^{M \times T}$ represents the measured body surface potentials. Here, $N$ denotes the number of nodes on the heart surface, $M$ the number of electrodes, and $T$ the number of temporal samples. \\\\
Due to the ill-posed nature of the inverse problem, directly solving this linear system with standard numerical methods results in unstable solutions. Therefore, regularization techniques must be applied to obtain a physiologically meaningful reconstruction, as discussed in the specialized literature \cite{hansen2010discrete}. In our numerical experiments, we employ zero-order Tikhonov regularization \cite{pullan2010inverse}, which balances the data-fitting term with constraints on the solution magnitude, ensuring stability and robustness:
\begin{equation}
    \hat{h} = \arg \min \left( \|Ah - g\|_2^2 + \lambda \|h\|_2^2 \right),  
\end{equation}    
where $\| \cdot \|_2$ denotes the $l_2$-norm, and $\lambda \in \mathbb{R}$ is the regularization parameter. The optimal value of $\lambda$ is determined using the L-curve criterion\cite{hansen1993use, molero2024improving}, a method widely used in ECGI to achieve a balance between solution stability and accuracy.

\subsection*{Volumetric reconstruction}
\begin{figure}[h]
    \centering
    \includegraphics[width=\linewidth, draft=false]{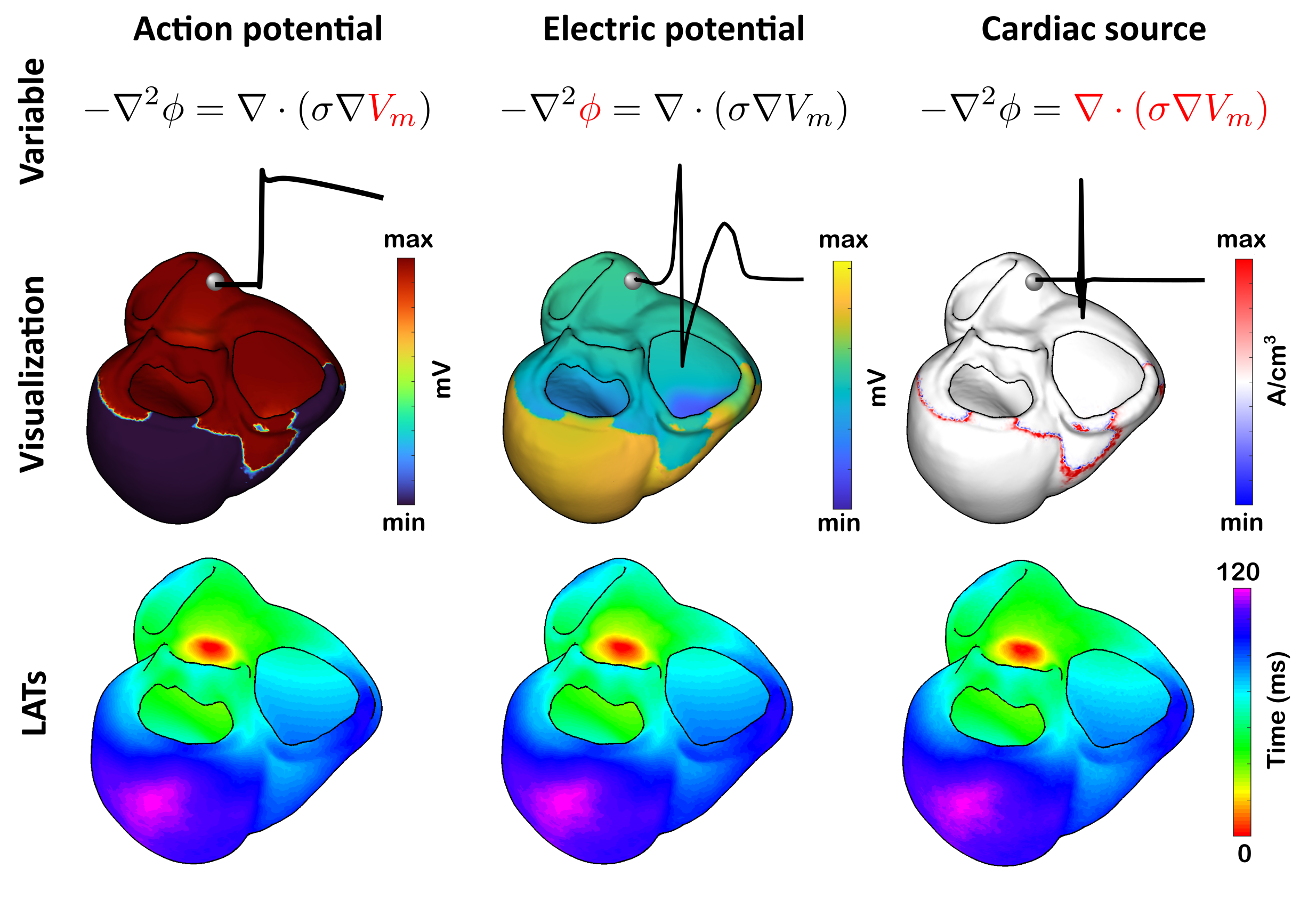}
    \caption{Comparison of local activation times (LATs) derived from different cardiac variables in a simulated PVC beat.    
   The top row presents the governing equation of electrical propagation in the body, with each highlighting the variable under analysis. Middle row shows spatial maps of the variable at a fixed time instant together with temporal traces at a fixed myocardial location. Bottom row shows the derived LAT maps.
   LATs are computed as the time of maximum upstroke for the action potential, and as the time of maximum positive rate of deflection for both the cardiac source and the electric potential. Although the variables differ in physical meaning, all provide consistent activation patterns across the myocardium.}
    \label{fig:heartVariables_LATs}
\end{figure}

We propose a volumetric ECGI formulation that reconstructs cardiac sources throughout the myocardial volume based on the mathematical framework of Poisson’s equation. By explicitly considering volumetric cardiac sources $f(x)$, this approach enables the estimation of three-dimensional activation patterns, overcoming the limitations of surface-based methods.\\\\
The torso is modelled as a homogeneous, quasi-static volume conductor with uniform conductivity to simplify the mathematical formulation of the volumetric ECGI problem. Based on established biophysical models \cite{malmivuo1995bioelectromagnetism, sundnes2007computing}, we formulate the inverse problem of reconstructing cardiac electrical sources from body surface potentials using the following equations:
\begin{align}
-\nabla^2 \phi(x) &= f(x), \quad x \in \Omega, \label{eq:SourceProblem_elliptic}\\
\phi(x) &= g(x), \quad x \in \Gamma, \\
\partial_{n_T} \phi(x) &= 0, \quad x \in \partial \Omega, \\
\int_{\Omega} f(x) \, dV &= 0,
\end{align}
where $\phi$ denotes the electric potential, $g$ the recorded or simulated body surface potentials, $f$ the cardiac sources that can only be non-zero in the heart, $\Omega$ the domain encompassing the body with boundary $\partial \Omega$ being the torso surface, $dV$ the standard Cartesian measure $d^3x$ on $\Omega$, and $\Gamma \subset \partial \Omega$ the electrode locations. The final equation ensures the cardiac sources satisfy the existence condition at each time instant.\\\\
Unlike the Cauchy problem (\ref{eq:CauchyProblem_elliptic}-\ref{eq:CauchyProblem_Neumann}),  the domain $\Omega$ in volumetric ECGI includes the heart, whereas the domain $\Omega_B$ in epicardial ECGI does not. Consequently, Poisson's equation \eqref{eq:SourceProblem_elliptic} replaces Laplace's equation \eqref{eq:CauchyProblem_elliptic} by incorporating the cardiac sources $f$ on the right-hand side. 
These cardiac sources can be derived from a simplified bidomain formulation coupled with a representation of the passive volume conductor between the heart and the torso surfaces~\cite{sundnes2007computing}. This leads to a concrete expression for the cardiac sources as a surrogate of the action potential:
\[
    f(x) = \begin{cases}
    \nabla \cdot (\sigma_i(x)\nabla V_m(x)), & x \in \Omega_H,\\
    0, & x \in \Omega_o.
    \end{cases}
\]
where $V_m$ denotes the action potential, $\sigma_i$ the intracellular conductivity tensor, $\Omega_H$ the heart volume and $\Omega_0$ the extramyocardial domain. This expression represents the net current source density (units: $A/cm^3$), which corresponds to the divergence of a current density modulated by the intracellular conductivity tensor. The formulation enables the estimation of a scalar field that, although not directly measurable, implicitly encodes the heart’s conductivity properties and the timing of cellular activation. By estimating this scalar field of sources and sinks, rather than the current density vector, our formulation simplifies the inverse problem while retaining essential electrophysiological information. Figure~\ref{fig:heartVariables_LATs} provides a first visual reference: the cardiac source is shown alongside the action potential and the extracellular electric potential, enabling side-by-side inspection of their shapes. \\

To implement the volumetric reconstruction, we discretize the domain $\Omega$ using a tetrahedral mesh generated from the triangulated surfaces of the torso and heart. The key contribution of this methodology is the use of Green’s functions under Neumann boundary conditions\cite{franklin2012green, duffy2015green}, which provide a mathematical framework for relating body surface potentials to volumetric cardiac sources, while respecting the intrinsic physical constraints of the electrophysiological system. \\\\
These functions define a linear system that connects electrode measurements to cardiac sources:
\begin{equation}
    Bf = g,
\end{equation}
where $B \in \mathbb{R}^{M \times P}$ is the transfer matrix, $f \in \mathbb{R}^{P \times T}$ represents the cardiac sources, and $g \in \mathbb{R}^{M \times T}$ represents the measured body surface potentials. Here, $P$ denotes the number of nodes of the heart volume, $M$ the number of electrodes, and $T$ the number of temporal samples. \\\\
Each element $B(i,j)$ of the transfer matrix $B$ is given by the Green's function $G(y_i, x_j)$, which are solutions to the partial differential equation:
\begin{align*}
    - \nabla^2 G(y_i, x) & = \delta_{y_i}(x), \hspace{0.25cm} x \in \Omega, \\
    \partial_n G(y_i, x) & = - \frac{1}{Area}, \hspace{0.25cm} x \in \partial \Omega,
\end{align*}
where $Area$ is the area of the torso surface, and $\delta_{y_i}$ is the Dirac delta function centered at the torso electrode location $y_i$. These functions capture how cardiac sources influence the potential measured on the torso surface. \\\\
Numerically, these relationships are computed using the finite element method (FEM), which solves Poisson’s equation within the discretized heart and torso geometries. Unlike the epicardial ECGI, where the transfer matrix $A$ maps torso potentials to epicardial potentials, the volumetric formulation extends this mapping with transfer matrix $B$ to the full cardiac volume. Consequently, the unknowns in the inverse problem are no longer epicardial potentials $h$, but rather the cardiac sources $f$.\\\\
As with the epicardial formulation, the ill-posed nature of the inverse problem necessitates regularization to prevent unstable solutions. We apply zero-order Tikhonov regularization \cite{pullan2010inverse} to ensure stability:
\begin{equation}
    \hat{f} = \arg \min_{M f = 0} \left( \|Bf - g\|_2^2 + \lambda \|f\|_2^2 \right),  
\end{equation}    
where $M$ enforces the existence condition of the cardiac sources, and $\lambda \in \mathbb{R}$ is determined using the L-curve criterion\cite{hansen1993use, molero2024improving}.

\subsection*{Local activation times}
In the final step of our methodology, the estimated cardiac sources are post-processed to compute local activation times (LATs). LATs can be extracted from various electrophysiological variables, including action potentials and electric potentials. In our formulation, the estimated cardiac sources act as a surrogate of the depolarization wavefront: they concentrate in a narrow band that sweeps through the myocardium, separating activated from resting tissue, and at any voxel produce a brief peak as activation arrives. Accordingly, we define the LAT as the time of the maximum positive rate of deflection (temporal slope) of the cardiac source signal, yielding a physiologically grounded, visually intuitive marker consistent with LATs derived from extracellular potentials and action-potential upstrokes (Fig.~\ref{fig:heartVariables_LATs}).
\\

To compute LATs from the estimated cardiac sources, we first apply a low-pass filter at each node of the heart tessellation to smooth temporal signals. Next, the activation time computation method\cite{invers2024regional} transforms each reconstructed signal into a sum of sinusoidal wavelets, weighting negative slope time samples proportionally to their slope amplitude. The activation time is then defined as the instant corresponding to the maximum amplitude of the transformed signal. \\\\
These LAT maps represent the temporal sequence of cardiac activation, which is crucial for replicating the electrophysiological dynamics of the heart. In epicardial ECGI, activation data is mapped onto the triangulation corresponding to the epicardium, restricting visualization to the outermost cardiac surface. In contrast, volumetric ECGI estimates activation data across the entire tetrahedralization, allowing for multiple visualizations of both epicardial, mid-myocardial and endocardial activation patterns. \\

To identify the origin of PVCs in ECGI reconstructions, using both epicardial and volumetric methods, we defined the site of earliest activation as the centroid of the subset of nodes with local activation times below the 10\textsuperscript{th} percentile. This approach provides a more robust estimate than selecting the single node with the earliest activation time.

\subsection*{Statistics and reproducibility}
All statistical analyses assess the performance of the proposed volumetric ECGI method. For simulations, we evaluate PVCs initiated at multiple sites, with replicates defined as independent initiation sites; performance is quantified by Euclidean and geodesic distances between the simulated and estimated origins of activation, and summarized across replicates using the median and interquartile range on the boxplots contained in Figure \ref{fig:metrics} (no formal hypothesis testing is performed given the exploratory scope and modest sample size). For clinical evaluation, we analyse four patients representing distinct pathologies; because of the small and heterogeneous cohort, agreement with electrophysiological reference information (clinically adjudicated diagnoses, electroanatomical maps, and fibrosis maps) is reported qualitatively. For the open myocardial infarction dataset, reconstructions are compared with the provided reference annotations to assess infarct-related abnormalities and are summarized descriptively.

\subsection*{Code availability}
The core implementation of the volumetric ECGI reconstruction was developed by Corify Care S.L. and are available from the corresponding author upon reasonable request.

\section*{Results}
\begin{figure}[h]
    \centering
    \includegraphics[width=\linewidth]{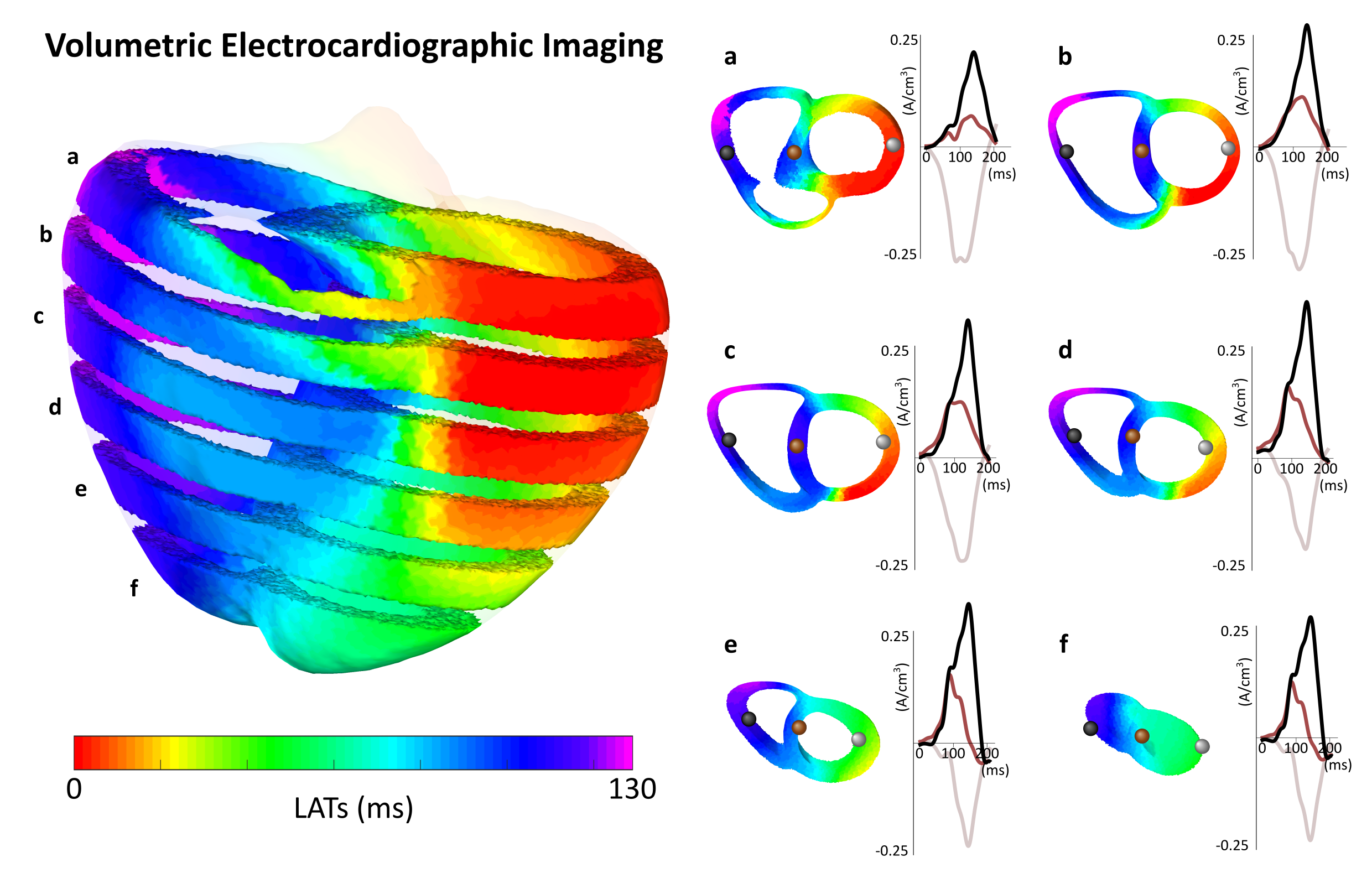}
    \caption{Volumetric ECGI activation sequence across multiple horizontal slices of the myocardium. The left panel illustrates the global activation sequence, visualized through horizontal slices, originating from a stimulation site on the left ventricular free wall. The right panel presents six individual slices (a–f), depicting activation propagation at different myocardial depths. Cardiac sources extracted from black, brown, and gray dots correspond to early, intermediate, and late activation times, respectively, showing distinct deflections that align with activation timing at varying depths within the myocardium.}
    \label{fig:visualizationVolumetric}
\end{figure}

We first illustrate the output of volumetric ECGI reconstruction in a patient undergoing monopolar stimulation of the left ventricular free wall, where the activation sequence is visualized across multiple horizontal slices (Figure~\ref{fig:visualizationVolumetric}). To evaluate the proposed algorithm, we perform a quantitative benchmark on sixteen simulated PVCs initiated at distinct ventricular regions (base, free wall and septum), comparing volumetric and classical surface-based ECGI using Euclidean and geodesic localization errors between the simulated earliest activation site and the estimated one (Figures~\ref{fig:simulations_ectopic_beats}, \ref{fig:metrics}). Finally, we evaluate clinical applicability in four representative patient cases, right ventricular outflow tract (RVOT) ectopic, left bundle branch block (LBBB), septal ventricular tachycardia (VT), and Wolff–Parkinson–White syndrome (WPW), as well as in two open post-myocardial infarction cases, assessing consistency with invasive or reference annotations (Figures~\ref{fig:patientPVC}-\ref{fig:patientWPW}). Together, these analyses probe visualization fidelity, localization accuracy, and clinical concordance across simulated and real-world scenarios. \\

To illustrate the output of the volumetric ECGI system, Figure \ref{fig:visualizationVolumetric} presents an example obtained from a patient undergoing monopolar stimulation in the left ventricular free wall. The activation sequence is visualized across multiple horizontal slices (a-f), providing a comprehensive representation of cardiac excitation. \\\\
The left panel of Figure \ref{fig:visualizationVolumetric} shows the activation pattern evolving over time, revealing the origin of activation in agreement with the monopolar stimulation site. A clear spatio-temporal coherence is observed across the epicardium, endocardium, and myocardium, demonstrating the system’s ability to resolve activation dynamics throughout the entire myocardial volume. In the right panel of Figure \ref{fig:visualizationVolumetric}, each slice highlights a distinct transmural gradient of activation, further confirming consistency with the expected physiological propagation. The reconstructed cardiac sources at different depths reveal a distinct deflection that aligns with the earliest activation sites across the slices, reinforcing the accuracy of the volumetric ECGI reconstruction.\\\\
Taken together, these findings show that volumetric ECGI enables a more comprehensive analysis of activation, capturing global conduction trends across the entire heart. This ability to resolve transmural activation patterns highlights its potential for improving non-invasive arrhythmia localization and guiding targeted therapies.

\subsection*{Comparative evaluation}
Premature ventricular contractions (PVCs) originating from sub-endocardial or mid-myocardial regions present a challenge for epicardial ECGI, as this approach primarily estimates epicardial potentials. To evaluate the accuracy of both surface and volumetric ECGI methodologies, a benchmark reference was developed using a dataset of simulated PVCs with known activation origins. Localization errors were quantified across different ventricular regions using Euclidean and Geodesic distances, which capture differences between estimated and simulated activation origins. These metrics provide clinical relevance by assessing the impact of each ECGI formulation on the accuracy of activation sites detection.\\

A set of 16 simulated PVCs was generated using anatomically detailed bidomain models in openCARP\cite{openCARP-paper, openCARP-sw}, incorporating realistic ventricular geometries and surrounding structures as described in Sánchez et al.\cite{sanchez2025enhancing}. The benchmark dataset included activations in different ventricular regions: six originating from the ventricular base, six from the septal wall , and four from the free wall. The origin of the ventricular beat was defined as the centroid of the subset of nodes with activation times below the $10^{th}$ percentile. Both surface and volumetric ECGI were evaluated using the same 128 surface potentials as input signals, providing a standardized basis for comparison. Further details on the numerical setup, mesh properties, and conductivity parameters are provided in the Methods section. \\

\begin{figure}[h]
    \centering
    \includegraphics[width=\linewidth]{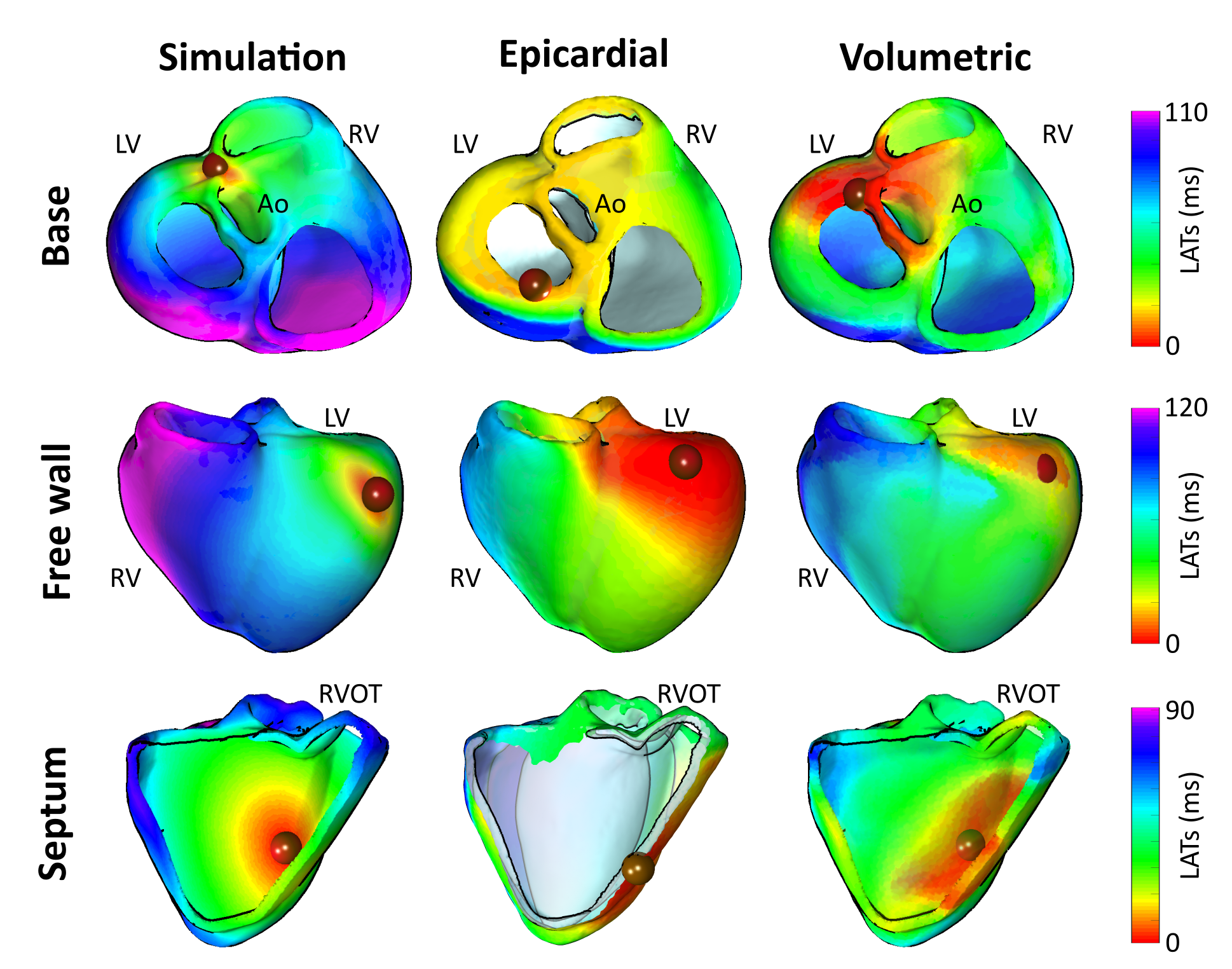}
    \caption{Comparison of LATs from the gold standard, epicardial and volumetric ECGI for ventricular ectopic beats localization. Activation maps reconstructed using different methodologies are presented in columns: gold standard simulations (left column), epicardial ECGI (middle column), and volumetric ECGI (right column). The rows correspond to three different simulations of ectopic beats originating from the ventricular base (top), free wall (middle), and septum (bottom). Red spheres mark the true and estimated sites of earliest activation. } 
    \label{fig:simulations_ectopic_beats}
\end{figure}

Figure \ref{fig:simulations_ectopic_beats} allows for a direct comparison of activation maps reconstructed by epicardial and volumetric ECGI for three distinct ectopic beats located in the ventricular base, free wall, and septum (in rows). \\
In the case of the ectopic beat located at the ventricular base (first row of Figure \ref{fig:simulations_ectopic_beats}), epicardial ECGI reconstruction exhibited spatial distortion, whereas the volumetric approach successfully avoided this distortion and allowed for a more accurate identification of the activation origin. This result highlights the volumetric method’s ability to handle anatomically realistic geometries, including the ventricular base with its distinct outflow tracts. For the free wall ectopic beat (second row of Figure \ref{fig:simulations_ectopic_beats}), both methods provided a similar activation sequence, leading to the same clinical interpretation. In contrast, for the septal ectopic beat (third row of Figure \ref{fig:simulations_ectopic_beats}), the epicardial ECGI reconstruction exhibited incomplete activation, missing crucial details within deeper myocardial layers as the location of the earliest activation site at the septal wall. In comparison, the volumetric approach successfully reconstructed a more physiologically consistent activation pattern throughout the septal wall.\\

\begin{figure}[h]
    \centering
    \includegraphics[width=\linewidth]{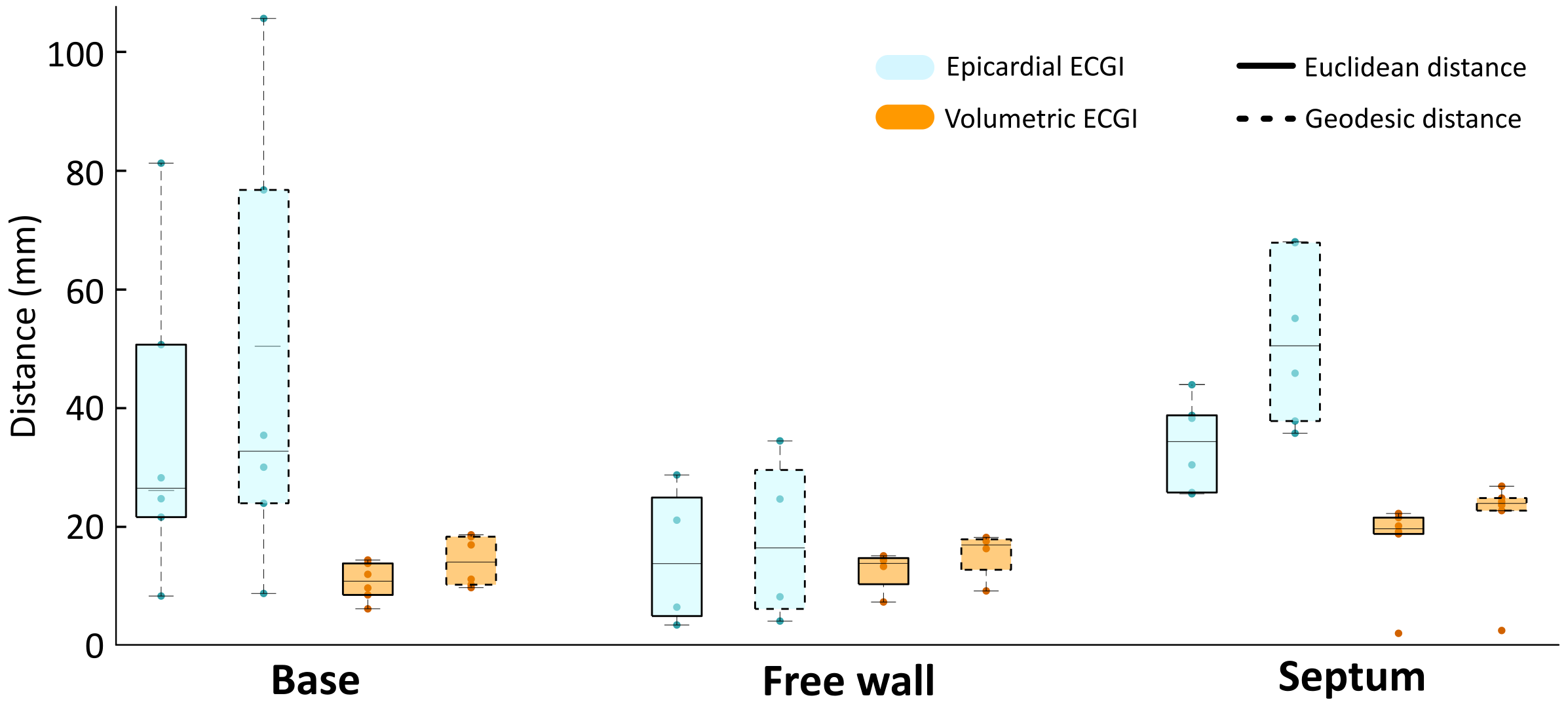}
    \caption{Localization errors in PVC detection using epicardial and volumetric ECGI. Boxplots compare Euclidean (solid lines) and geodesic (dashed lines) distance errors for PVC localization across three ventricular regions: base, free wall, and septum. Results are shown for epicardial ECGI (blue) and volumetric ECGI (orange).}
    \label{fig:metrics}
\end{figure}

The comparison between epicardial and volumetric ECGI revealed substantial differences in localization accuracy, particularly in deep myocardial regions (Figure \ref{fig:metrics}). The analysis on the entire simulation set showed that LATs estimated with the epicardial ECGI resulted in an average Euclidean distance of 29.82$\pm$18.87 mm and geodesic distance of 41.28$\pm$27.52 mm between actual and estimated earliest activation sites, while volumetric ECGI reduced these errors down to 13.53$\pm$5.85 mm and 16.82$\pm$6.80 mm, respectively. This corresponds to a mean global error reduction of 54.6$\%$ for the Euclidean distance and 59.3$\%$ for the geodesic distance, underscoring a substantial improvement in localization accuracy. \\
Regional analysis of the localization errors revealed that PVCs originating from the septal wall exhibited the highest discrepancies, with epicardial ECGI yielding an average Euclidean error of 33.79$\pm$7.61 mm and geodesic error of 51.62$\pm$14.25 mm. Volumetric ECGI improved these estimations to 17.20$\pm$7.57 mm and 20.65$\pm$9.03 mm, respectively. For PVCs located in the ventricular base, epicardial ECGI resulted in Euclidean errors of 35.87$\pm$26.10 mm and geodesic errors of 46.65$\pm$36.61 mm, while volumetric ECGI reduced these values to 10.62$\pm$3.20 mm and 14.02$\pm$4.21 mm. In the free wall region, where differences between the two approaches were smaller, epicardial ECGI yielded Euclidean errors of 14.78$\pm$11.97 mm and geodesic errors of 17.70$\pm$14.16 mm, while volumetric ECGI showed a slightly improved localization with errors of 12.38$\pm$3.53 mm and 15.26$\pm$4.15 mm.\\

The consistent error reduction across all ventricular regions demonstrates the robustness of volumetric ECGI in improving localization. This is particularly evident for PVCs originating from deep myocardial structures such as the septum and anatomically complex geometries like the ventricular base. In these regions, the passive conductor assumption in epicardial ECGI contributes to higher localization errors, particularly due to its limitations in accurately accounting for the effects of concavities and convexities in the cardiac geometry, which can distort the reconstruction process. Notably, volumetric ECGI reduced geodesic distance by 69.9\% in the ventricular base and 60.0\% in the septal wall, highlighting its potential to improve arrhythmia localization in regions where traditional ECGI is less accurate. These findings highlight the impact of the inverse problem formulation on reconstruction accuracy, suggesting that a more accurate localization could improve the identification of ectopic foci in clinical applications and aid in tailoring patient-specific ablation planning.

\subsection*{Clinical applications}
Following quantitative evaluation by using computer models, volumetric ECGI was applied in clinical cases to assess its utility in mapping activation patterns in real-world scenarios. Four patients with distinct electrophysiological conditions were analyzed, each representing a different mapping challenge: a case of premature ventricular contractions (PVCs) in the right ventricular outflow tract (RVOT), a case of left bundle branch block (LBBB), {a case of septal ventricular tachycardia (VT), and a case of Wolff-Parkinson-White (WPW) syndrome. To further support comparative analysis, we also included two open-source cases with documented myocardial infarction from publicly available datasets. \\

\begin{figure}[h]
    \centering
    \includegraphics[width=\linewidth]{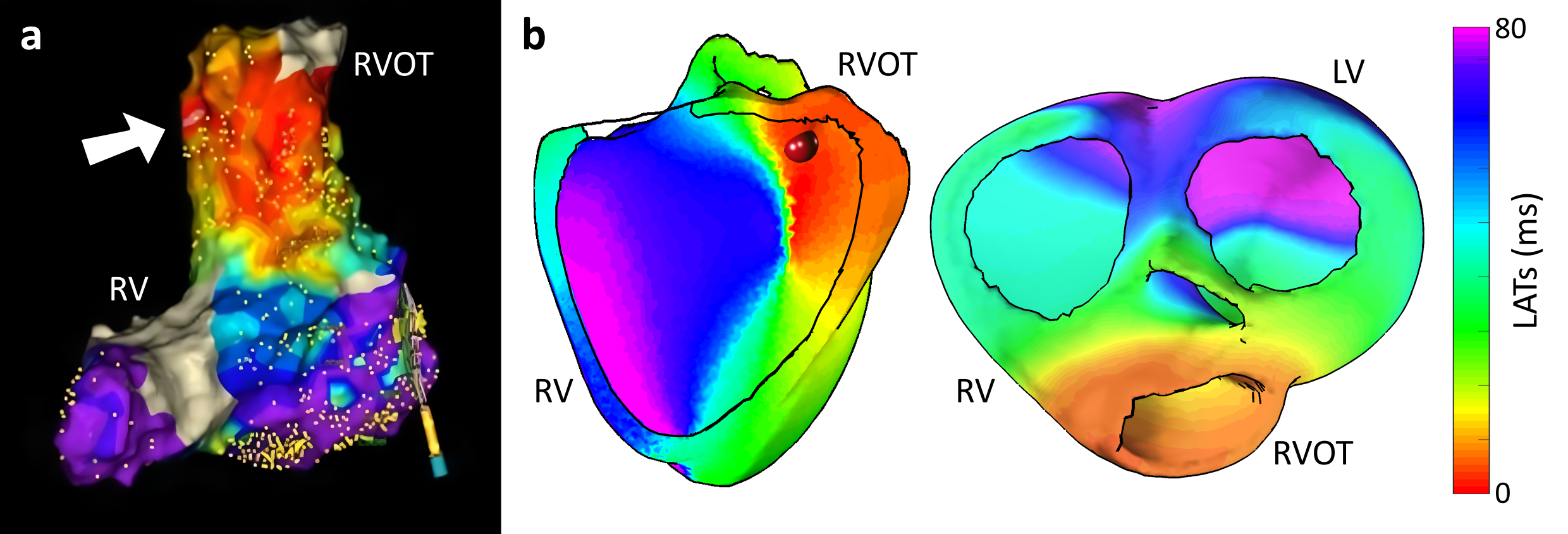}
    \caption{Volumetric ECGI activation maps of a patient with PVCs originating from the right ventricular outflow tract (RVOT). (a) Invasive electroanatomical mapping of the endocardial surface, with the earliest activation site highlighted. (b) Volumetric ECGI-reconstructed LATs map, showing the propagation of electrical activation from the RVOT through the right ventricular septal wall and base. A white arrow indicates the site of earliest activation identified by the electroanatomical map, while a red sphere marks the corresponding estimated origin in the reconstructed propagation.} 
    \label{fig:patientPVC}
\end{figure}

First, we analyzed a 62-year-old patient with symptomatic PVCs to assess the ability of volumetric ECGI to accurately localize the ectopic focus within the complex RVOT region. Figure \ref{fig:patientPVC}a displays the invasive electroanatomical map obtained via catheter-based mapping, confirming the RVOT origin. The volumetric ECGI activation map (Figure \ref{fig:patientPVC}b) successfully identified this ectopic origin, tracking the propagation sequence and spatial distribution of the activation. The reconstruction demonstrated the capability to resolve activation patterns with high spatial precision in this anatomically complex region, accurately distinguishing the PVC origin from adjacent structures. 

\begin{figure}[h]
    \centering
    \includegraphics[width=\linewidth]{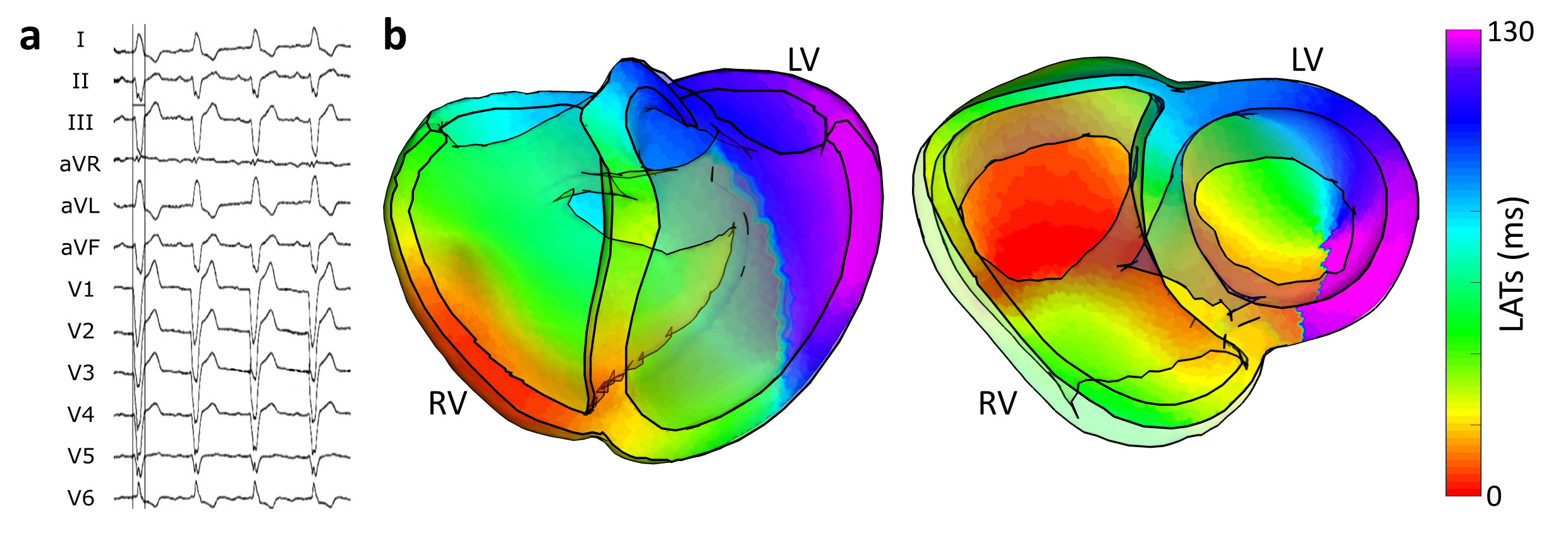}
    \caption{Volumetric ECGI activation maps of a patient with left bundle branch block (LBBB). (a) Standard 12-lead ECG confirming the LBBB diagnosis. (b) LATs map displaying delayed activation across the left ventricular myocardium, consistent with LBBB pathology. }
    \label{fig:patientLBBB}
\end{figure}

Next, the volumetric ECGI capability to map conduction abnormalities was evaluated in an 80-year-old patient diagnosed with LBBB. The clinical diagnosis was confirmed via 12-lead ECG (Figure \ref{fig:patientLBBB}a). Volumetric ECGI was applied to map the activation sequence (Figure \ref{fig:patientLBBB}b). The reconstruction revealed delayed activation of the left ventricular myocardium, with both endocardial and myocardial estimates displaying dyssynchronous conduction patterns consistent with previously reported LBBB characteristics.\\

Next, we evaluated volumetric ECGI’s capacity to map reentrant conduction in a 51-year-old patient presenting with ventricular tachycardia (VT). The arrhythmogenic substrate was characterized using the ADAS 3D fibrosis detection algorithm (Figure \ref{fig:patientVT}a), which identified septo-basal fibrotic substrates of the left ventricle. Volumetric ECGI was then applied to reconstruct the full three-dimensional activation sequence during VT (Figure \ref{fig:patientVT}b). The resulting activation map revealed reentrant activity traversing the channel between the identified septo-basal fibrotic substrates. This case highlights the potential of volumetric ECGI to resolve intramural reentrant circuits in complex clinical scenarios. \\
\begin{figure}[h]
    \centering
    \includegraphics[width=\linewidth]{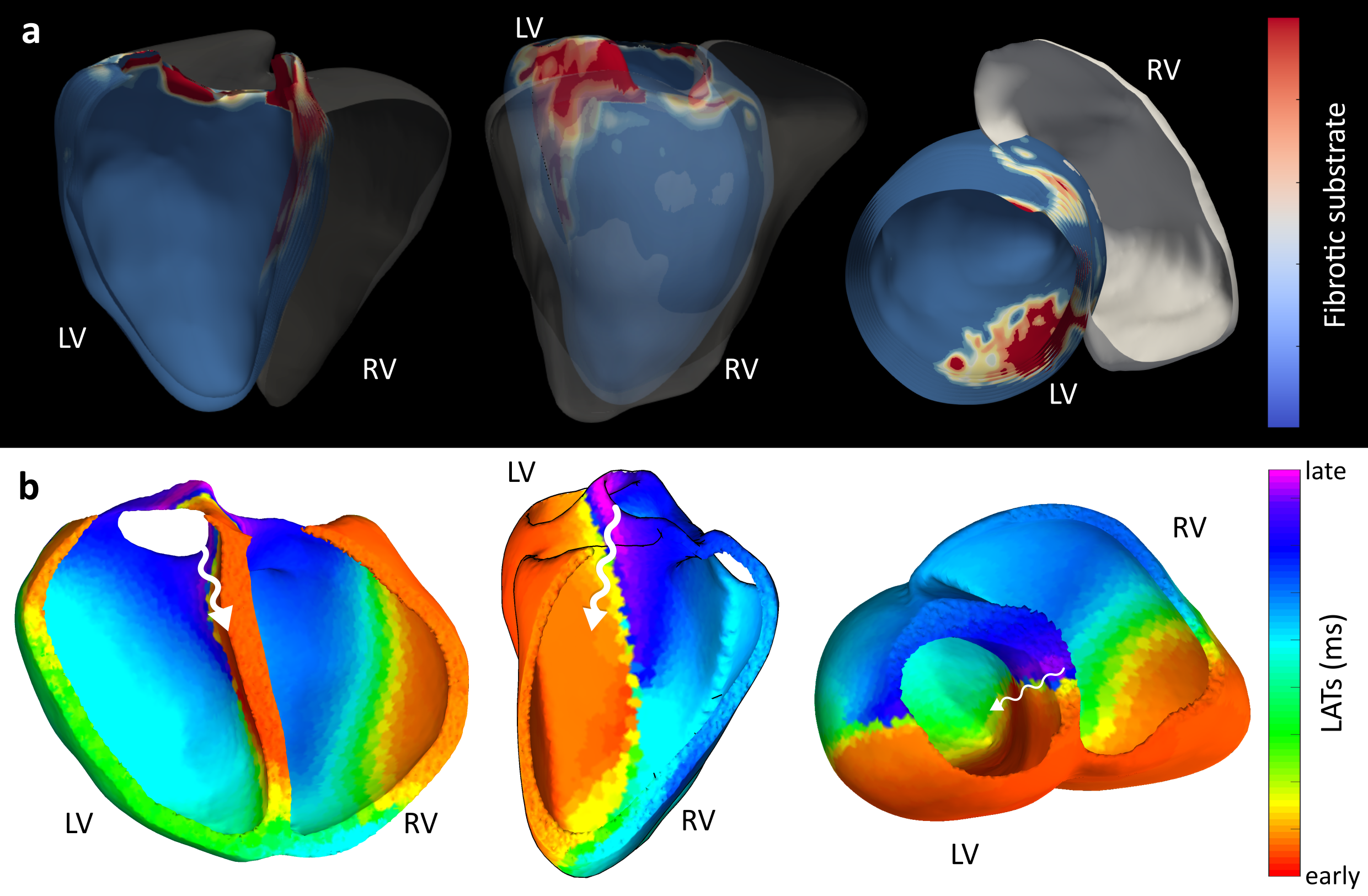}
    \caption{Volumetric ECGI activation maps of a patient with septal ventricular tachycardia (VT). 
    (a) Three-dimensional fibrosis map generated by the ADAS 3D algorithm, highlighting a fibrosis scar on the upper part of the septal wall.
    (b) Volumetric ECGI LAT map depicting a stable reentrant activation circuit confined to the septal wall, consistent with the VT mechanism. } 
    \label{fig:patientVT}
\end{figure}

Finally, using the multi-chamber capability of our approach, volumetric ECGI was applied to a 58-year-old patient diagnosed with WPW syndrome. Invasive catheter mapping confirmed the presence of an accessory pathway in the postero-septal region (Figure \ref{fig:patientWPW}a). The volumetric ECGI reconstruction (Figure \ref{fig:patientWPW}b) successfully identified the propagation of electrical activity across both the atria and ventricles. This allowed for accurate identification of the conduction pathway and its propagation sequence, closely matching the invasive reference and offering a comprehensive assessment by integrating atrial and ventricular activation in a single solution. \\

\begin{figure}[h]
    \centering
    \includegraphics[width=\linewidth]{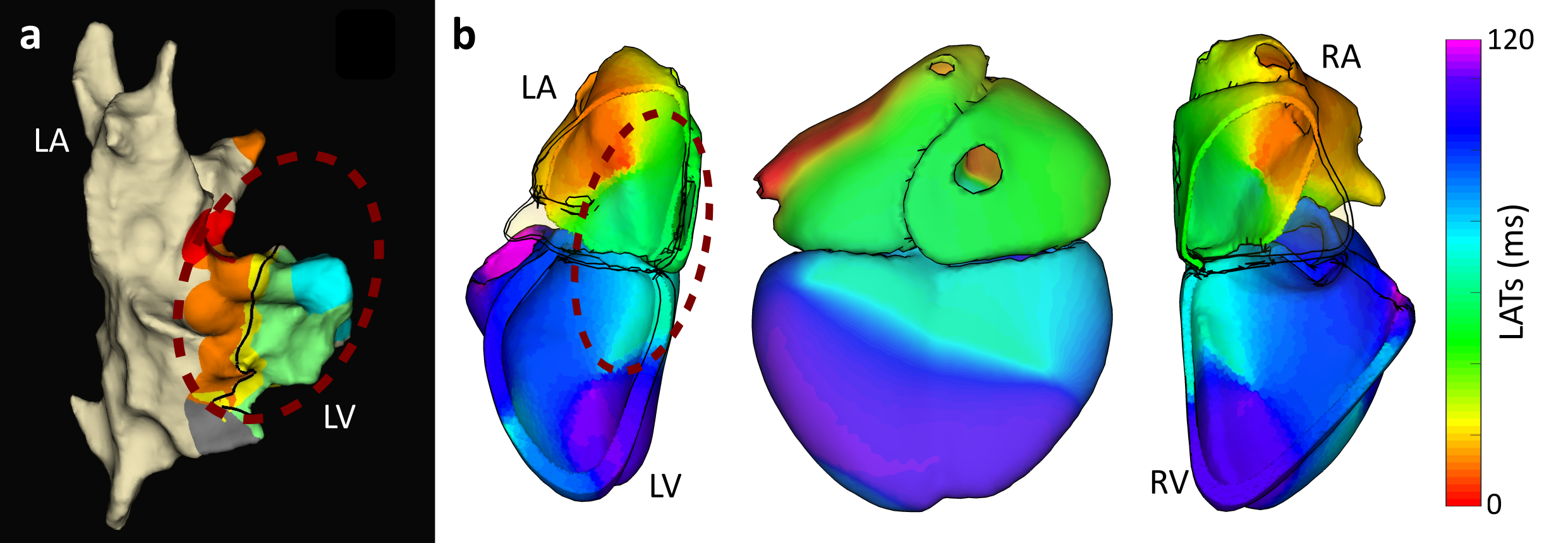}
    \caption{Volumetric ECGI activation maps of a patient with Wolff-Parkinson-White (WPW) syndrome. (a) Invasive electroanatomical mapping of the endocardial surface, highlighting the earliest activation site at the location of the accessory pathway (dashed red circle). (b) Three different views, left ventricular, posterior, and right ventricular perspectives, illustrating activation propagation across the septal wall.}
    \label{fig:patientWPW}
\end{figure}

To support comparative analysis, we employed experimental datasets from two post-myocardial infarction patients provided by the 2007 Computers in Cardiology Challenge\cite{goldberger2000physiobank}. \\
To evaluate the performance of the method, we employ standard metrics commonly used in myocardial infarction assessment: location, centre, extent, and segment overlap. The infarct location is identified based on the spatial distribution of abnormal regions within the myocardium; the centre refers to the region containing the centroid of the infarcted area; the extent quantifies the proportion of affected myocardial tissue; and the segment overlap measures the agreement between the estimated and reference infarcted segments, based on the standard 17-segment model of the left ventricle\cite{american2002standardized}.\\
To identify the segments labelled as fibrotic tissue, we considered the Total Activation Times (TATs) within each region \cite{reventos2025non}. Each metric captures a distinct aspect of infarct characterization, enabling a comprehensive comparison with the reference annotations. \\\\
Case 3 corresponds to a large infarct involving the inferior-septal, inferior, and inferolateral regions (AHA segments 3, 4, 5, 9, 10, 11, 12, 15, and 16). The infarct centroid is located between segments 10 and 11, with an overall extent of 52\%. Our method reconstructed a fibrotic substrate encompassing segments 4 to 7, 9 to 14, and 16, with the estimated centroid located in segment 11. The estimated infarct extent was 64.3\%, and the segment overlap was 0.65. \\\\
Case 4 represents a dual infarct involving the mid-apical inferior and basal anterior regions (AHA segments 1, 9, 10, 11, 15, and 17). The infarct centroid is located in segment 15, with an overall extent of 14\%. Our method identified segments 1, 3, 4, 6, and 9 to 14 as fibrotic substrate. The estimated centroid corresponded to segment 9, with an infarct extent of 24.5\% and a segment overlap of 0.49. \\

\section*{Discussion} 
This study demonstrated the feasibility of volumetric ECGI in reconstructing cardiac activation sequences across different electrophysiological conditions, improving localization accuracy in both simulated and clinical cases. Unlike classical ECGI approaches, our method explicitly allows for an estimation of cardiac sources from Poisson's equation, using Green's functions to relate them to body surface potentials, enabling a volumetric reconstruction of myocardial activation. By estimating activation throughout the myocardial volume rather than relying solely on epicardial surface potentials, volumetric ECGI provided a more comprehensive representation of conduction patterns.\\\\ 

\subsection*{Performance in volumetric cardiac mapping} 
In simulated scenarios, volumetric ECGI consistently reduced localization errors across different ventricular regions, particularly in septal and base PVCs, where traditional ECGI approaches exhibit higher errors due to their reliance on epicardial potentials estimation. Our findings align with these observations, showing that epicardial ECGI resulted in significantly larger geodesic errors in these locations, while volumetric ECGI improved precision.\\\\ 
The advantages of volumetric ECGI were further confirmed in clinical cases. In the PVC patient, the activation map localized the ectopic focus within the RVOT, a region where precise non-invasive mapping is difficult yet crucial for planning ablation, especially when invasive procedures are scarce\cite{parreira2022assessment}. In the LBBB case, volumetric ECGI accurately reconstructed the expected dyssynchronous activation pattern, offering detailed transmural insights valuable for CRT patient selection beyond standard ECG\cite{ploux2013noninvasive}. Similarly, in the VT case, volumetric ECGI precisely mapped the reentrant activity induced by the septo-basal substrate. This substrate was identified by the ADAS 3D fibrosis detection algorithm and localized to the septal wall. These results demonstrate how volumetric ECGI overcomes the classical epicardial-only limitation of conventional ECGI approaches~\cite{sapp2020mapping}. In the WPW patient, volumetric ECGI successfully estimated atrial and ventricular activation, overcoming the single-chamber limitation of conventional ECGI (including the difficulty in estimating septal pathways in WPW) and providing comprehensive pathway characterization essential for pre-ablation planning\cite{ghosh2008cardiac}. Finally, in the myocardial infarction cases, volumetric ECGI successfully identified the infarcted region with reasonable accuracy in case 3, although a larger spatial displacement was observed in case 4.\\\\ 

Volumetric ECGI thus consistently demonstrated improved localization performance across both simulated and clinical settings, particularly in anatomically complex regions such as the septum and ventricular base. This is in line with prior research showing that conventional ECGI struggles to accurately reconstruct activations originating beyond the epicardium, resulting in significant localization errors\cite{schuler2021reducing, van2023basis}.\\\\
ECGI has shown significant challenges in localizing PVCs originating from the septum and ventricular base. In particular, traditional ECGI based on electric potential estimation is known to provide unreliable reconstructions beyond the epicardium, leading to substantial localization errors in these regions, as observed in PVC simulations\cite{schuler2021reducing, van2023basis, duchateau2019performance}. Consistent with these findings, our study demonstrated that traditional epicardial ECGI exhibited considerable localization inaccuracies in septal and base PVCs, whereas volumetric ECGI significantly improved localization precision. For instance, in septal PVCs, the Euclidean distance error was reduced from 33.79$\pm$7.61 mm with epicardial ECGI to 17.20$\pm$7.57 mm with volumetric ECGI, highlighting the potential of volumetric reconstructions to enhance source estimation in complex anatomical regions. \\\\ 
Previous studies have shown that ECGI can enhance the localization accuracy of outflow tract ventricular arrhythmias in patients with PVCs compared to standard ECG methods\cite{zhou2020comparative}. Our results further demonstrate that volumetric ECGI accurately identifies PVC origins within the outflow tracts, even in anatomically complex and realistic geometries. This improved spatial resolution and localization accuracy provides a valuable tool for guiding catheter ablation, potentially reducing intervention times and improving procedural outcomes. \\\\ 
Studies on LBBB have demonstrated that ECGI can effectively capture global conduction delays and quantify ventricular electrical uncoupling (VEU), a key predictor of response to cardiac resynchronization therapy (CRT)\cite{ploux2013noninvasive}. Our results confirm that volumetric ECGI accurately reconstructed the dyssynchronous activation patterns characteristic of LBBB, as diagnosed by the 12-lead ECG. The integration of volumetric ECGI could further refine patient selection for CRT, as prior research\cite{ploux2013noninvasive} suggests that improved spatial resolution in ECGI is associated with better CRT outcomes.\\\\ 

Previous ECGI studies in VT have demonstrated its ability to non-invasively localize arrhythmia mechanisms, overcoming the interpretive limitations of standard ECG mapping in both hemodynamically stable and unstable VT~\cite{graham2020evaluation}. In this work, we present a clinical case in which volumetric ECGI achieved a single-beat reconstruction of a septal reentrant circuit, in concordance with the fibrosis substrate identified by the ADAS 3D algorithm. Our observation is consistent with a growing clinical literature that reports intramural substrate/driver estimation from body-surface data using ECGI formulations beyond epicardial potentials~\cite{tsyganov2018mapping,wang2018non,schulze2017ecg}. These studies illustrate feasibility but remain limited by small cohorts and predominantly qualitative or indirect validation (e.g., concordance with scar imaging or pace-map surrogates), so they do not yet establish robust intramural localization accuracy. Such volumetric mapping of intramural VT substrates, previously inaccessible to epicardial only ECGI, holds significant promise for guiding non-invasive ablation therapies, including radioablation\cite{cuculich2017noninvasive,parreira2025defining}. \\\\ 

Previous studies have demonstrated that ECGI can noninvasively localize accessory pathways in WPW syndrome\cite{ghosh2008cardiac}. However, traditional methods primarily focus on ventricular activation, often overlooking the role of atrial activity in pathway localization. Our study expands on these findings by incorporating volumetric ECGI reconstructions of both atria and ventricles, allowing for an independent assessment of the P and QRS waves. This multi-chamber approach provides a more comprehensive characterization of WPW dynamics, as the accessory pathway influences both atrial and ventricular depolarization patterns. By incorporating both atrial and ventricular reconstructions, our methodology improves preprocedural planning for catheter ablation, potentially improving procedural efficiency and patient outcomes\cite{ghosh2008cardiac}. \\\\
Previous studies on myocardial infarction detection using ECGI have demonstrated the potential of this technology to identify fibrotic tissue based on estimated electrical activity~\cite{wang2009physiological, wang2010noninvasive, xu2014noninvasive}.
In this work, we analysed two clinical cases originally included in the 2007 PhysioNet Challenge. In case 3, we successfully identified the correct centre of the fibrotic region, although the estimated extent was slightly larger and the location mildly displaced.
Wang et al. reported comparable accuracy, correctly identifying the centroid segment with a segment overlap of $66.7\%$\cite{wang2010noninvasive, xu2014noninvasive}, while other studies achieved improved overlap values up to $90\%$\cite{wang2009physiological}.
In case 4, the fibrotic region was overestimated and displaced, resulting in a less accurate reconstruction. Previous studies also struggled to accurately recover the fibrotic region in this case, reporting centroid displacement and a segment overlap of only $33.3\%$\cite{wang2010noninvasive}. In contrast, other methods achieved better agreement, with both the centroid and segment overlap reaching $75\%$\cite{xu2014noninvasive}.\\
These comparisons indicate that, although our methodology was designed generically and not specifically optimized for infarct detection, it achieved reasonable accuracy. Nonetheless, the use of more targeted regularization strategies could significantly improve the delineation of infarcted regions. \\\\ 
These results establish this technique of Volumetric Source Imaging as a method capable of improving localization accuracy, particularly in regions that are problematic for epicardial-based methods. This enhanced diagnostic capability, especially when coupled with imageless approaches, could improve pre-procedural planning and clinical decision-making, potentially increasing access to effective arrhythmia management in diverse settings, including resource-constrained environments.

\subsection*{Methodological comparison with prior work}
To contextualize the results obtained with the proposed methodology, it is important to compare it with prior research on ECGI. Several studies have highlighted how ECGI performance is influenced by the underlying source model, particularly in cases involving deep myocardial structures\cite{van2014comparison,schuler2021reducing, van2023basis, dogrusoz2023comparison, messnarz2004comparison}. Hence, it is necessary to compare our approach alongside other methodologies specifically developed to overcome the classical epicardial limitations of ECGI.\\

Traditional ECGI models have focused on epicardial potential estimation, limiting their ability to accurately localize intramural or endocardial sources. To overcome these limitations, various techniques have been developed, ranging from dipole-based reconstructions\cite{dogrusoz2023comparison, ondrusova2023two} to volumetric estimation approaches\cite{he2002noninvasive, nielsen2007possibility, diallo2021volume}. Our approach, a technique of Volumetric Source Imaging, differs from traditional ECGI methods by employing cardiac source estimation based on Green’s functions, which establishes a computationally efficient and physiologically meaningful relationship between electrode potentials and cardiac sources. \\\\
To address the limitations of traditional ECGI, various methods have been developed to extend the estimation of cardiac activity beyond the epicardium. One of the most widely adopted approaches is the estimation of endocardial and epicardial action potentials using the Equivalent Double Layer (EDL) model~\cite{geselowitz1992description}. This model consists of a dipole layer distributed across the myocardial surface, where the action potentials are defined, and it has been shown to accurately simulate electrograms and body surface potentials, as demonstrated in platforms like ECGSIM~\cite{van2004ecgsim}. Based on this model, several studies have investigated non-linear estimation techniques to reconstruct action potentials from body surface potentials~\cite{huiskamp2002depolarization}. Linear estimation approaches have also been proposed. In particular, Schuler et al.\cite{schuler2019delay} compared traditional epicardial-based ECGI with transmembrane voltage estimation on an endo-epicardial surface and found that the latter improved septal activation accuracy in PVC localization. Similarly, Dogrusoz\cite{dogrusoz2023comparison} studied the dipole-based source model in comparison with traditional ECGI, further highlighting how the choice of cardiac source model influences localization precision.\\\\ 
Several studies over the past two decades have explored volumetric ECGI to estimate cardiac activation throughout the myocardium. Early works by He et al. \cite{he2002noninvasive, he2003noninvasive} employed neural networks to localize premature ventricular contraction (PVC) origins, and later introduced the estimation of equivalent current density as an alternative source model \cite{liu2006noninvasive, yu2015temporal}. Unlike our approach, these methods reconstruct a three-dimensional vector field, from which local activation times (LATs) are derived as the instants of maximum amplitude \cite{liu2006noninvasive, yu2015temporal}. In contrast, our method estimates a scalar surrogate of the transmembrane current, providing a direct proxy for LATs without requiring the reconstruction of full vector fields.\\

A separate class of methods focuses on estimating transmembrane action potentials within the myocardium. Wang et al. investigated this problem using various types of sophisticated regularization methods, including stochastic state-space formulations with unscented Kalman filtering~\cite{wang2009physiological}, total variation regularization~\cite{xu2014noninvasive}, and Bayesian inference to jointly estimate the solution and errors in the prior model~\cite{ghimire2019noninvasive}, among others. Nielsen \cite{nielsen2007possibility} also proposed a one-shot inverse method incorporating physiological priors to estimate action potentials. While these methods provide highly detailed reconstructions of the cardiac state, they often require advanced computational frameworks. Our approach, in contrast, aligns with the goal of providing an accessible volumetric solution, prioritizing a balance between clinical utility and computational efficiency by estimating a direct surrogate for activation times. \\
More recently, Diallo et al. \cite{diallo2021volume} developed a linear inverse source model using a variational formulation and the finite element method to derive the forward operator. In their approach, the estimated cardiac sources are post-processed via an additional forward simulation to compute the electric potentials, from which LATs are derived. This two-step process introduces both computational overhead and potential error accumulation. In contrast, our method derives the forward operator using Green’s functions under Neumann boundary conditions, and computes LATs directly from the reconstructed source term, eliminating the need for a second forward simulation. This simplification reduces both complexity and computational cost. \\

Overall, our formulation offers a complementary approach to existing ECGI methods that seek to overcome epicardial limitations, providing an accessible, fast, and robust framework that retains physiological relevance while enhancing computational feasibility. This balance makes it particularly well-suited for integration into clinical pipelines, where robustness, interpretability, and speed are essential. \\

\subsection*{Limitations and future directions}
This study presents a practical approach to volumetric ECGI, offering accurate activation mapping through a computationally efficient and clinically accessible framework. While the results obtained in both simulated and clinical settings underscore its potential, several important limitations remain. Addressing these will be critical for improving the physiological fidelity, robustness, and generalizability of the method across diverse clinical contexts. \\

Volumetric inverse problems inherently lack uniqueness, as multiple internal source configurations can produce indistinguishable body surface potentials, making the inverse problem severely ill-posed. To stabilize the inversion, we employ zero-order Tikhonov regularization \cite{hansen2010discrete}, which suppresses high-frequency components while preserving the spatial structure of the cardiac sources.
Despite its simplicity, this method proves sufficient for the intended clinical utility of our approach. Since the objective of this work is to obtain volumetric LAT maps in a simple and computationally efficient manner, the limited resolution of the estimated wavefronts does not significantly affect the final outcome. However, more accurate reconstructions of cardiac sources would enable more physiologically meaningful estimations and support the computation of other electrophysiological biomarkers of clinical interest. Therefore, future extensions could incorporate anatomical priors, temporal regularization \cite{schuler2019delay}, or sparsity constraints~\cite{rahimi2013lp} to improve physiological accuracy and robustness. \\\\ 

While volumetric ECGI has demonstrated strong performance in both simulated and clinical cases, broader validation is needed across diverse patient populations with varying cardiac geometries and arrhythmogenic substrates. Prior research\cite{minchole2019mri} has highlighted the impact of anatomical variability on ECG signal interpretation, emphasizing the need for further studies in heterogeneous cohorts. \\\\
Despite its advantages, volumetric ECGI remains sensitive to geometric and conductivity assumptions. Keller et al.\cite{keller2010ranking} noted that ECGI reconstructions rely heavily on accurate conductivity modelling, which directly affects source localization precision. 
In this work, the conductivity was assumed to be homogeneous throughout the entire domain to adapt for a clinical scenario in which the true conductivity values are unknown. We acknowledge that this simplification introduces systematic errors that can be observed in the reconstructions, for example, in Figure~\ref{fig:visualizationVolumetric}, the estimated cardiac source on the septal wall exhibits a smaller amplitude compared to the one estimated on the free wall. Although this represents a clear limitation, it does not significantly affect the objective of the present study. Since the shape of the estimated signals remains consistent, the resulting LATs are only minimally impacted. Future work should focus on incorporating spatially varying conductivities to enhance volumetric ECGI accuracy and reduce these dependencies.  \\\\
Additionally, our findings suggest that signal-to-noise ratio (SNR) impacts reconstruction accuracy, warranting further investigation. The influence of electrode count on PVC localization should also be explored, as prior work\cite{ondrusova2023two} has reported increased localization errors with reduced electrode density. \\\\

The results presented in this study highlight the potential of volumetric ECGI as a valuable tool for clinical decision-making in electrophysiology. Specifically, our findings suggest that volumetric ECGI could enhance preprocedural planning by improving localization accuracy in anatomically complex regions, such as the septum and ventricular base, where conventional ECGI often struggles. By providing a more comprehensive representation of cardiac activation, volumetric ECGI could assist electrophysiologists in identifying arrhythmogenic substrates more precisely, reducing reliance on invasive mapping and improving treatment outcomes.\\\\ 
Prospective studies have demonstrated that ECGI-based mapping significantly improves arrhythmia localization, leading to faster and more effective ablation procedures. In a randomized trial, Erkapic et al.\cite{erkapic2015clinical}  showed that ECGI-guided mapping increased localization accuracy to 95.2\%, compared to 38.1\% with 12-lead ECG algorithms, and reduced procedural time by 30\%. Given these findings, integrating volumetric ECGI into clinical workflows could further optimize ablation strategies, particularly for arrhythmias originating in deep myocardial structures, where traditional approaches are less reliable. 
Building on this, volumetric ECGI can be integrated into noninvasive ablation pipelines: Parreira reported a case of a septal VT unsuitable for catheter ablation where the treatment target was defined by combining ECGI activation maps with CT-derived scar geometry, enabling successful stereotactic radioablation of the septal circuit~\cite{parreira2025defining}. 
Beyond ablation, volumetric ECGI may also refine patient selection for CRT and improve therapy outcomes. Our study supports the potential of volumetric ECGI for CRT optimization by demonstrating that volumetric ECGI can reconstruct global and regional activation delays, offering a more comprehensive assessment of electrical dyssynchrony. Given these findings, integrating volumetric ECGI into clinical workflows could result in more efficient and targeted ablations, particularly in cases where conventional ECGI has shown limitations. \\\\ 

Finally, further clinical validation should focus on pathology-specific prospective trials and explicitly quantify performance against invasive ground truth. We acknowledge that intramural localization of activation origins and re-entry circuits remains an open challenge; accordingly, we avoid claims of definitive intramural localization in the absence of appropriate quantitative error metrics and invasive reference standards. In PVC or VT ablation candidates, volumetric ECGI should be systematically compared with standard ECGI and electroanatomical mapping to report spatial localization errors and diagnostic performance. In CRT candidates, evaluation should emphasize whether volumetric ECGI provides incremental predictive value for response by correlating non-invasive activation delays with invasive measures and clinical outcomes. Because our current clinical cases rely primarily on qualitative visual comparisons, we will conduct rigorous quantitative studies in larger patient cohorts to support the future positioning of volumetric ECGI as a routine tool for non-invasive arrhythmia mapping and therapy guidance.\\

\subsection*{Conclusions}
A volumetric ECGI methodology has been presented for three-dimensional non-invasive cardiac mapping. By explicitly estimating cardiac sources from Poisson’s equation and relating them to body surface potentials through Green’s functions, our approach enables a volumetric reconstruction of myocardial activation. This addresses the limitations of traditional ECGI, which is restricted to epicardial reconstructions, improving localization accuracy, particularly in deep or complex anatomical regions. \\\\
Validation across simulated and clinical cases underscores the agreement between volumetric ECGI activation maps and invasive electroanatomical mapping. These findings highlight the potential clinical applications of volumetric ECGI in preprocedural planning, ablation guidance, and CRT patient selection. While further validation in larger patient cohorts is required, volumetric ECGI represents a promising advancement in non-invasive cardiac electrophysiology, offering improved diagnostic precision and treatment planning.

\section*{Data availability}
We adhere to the Scientific Reports data availability policy. The datasets generated and/or analysed during the current study are available from the corresponding author upon reasonable request. The source codes used to generate the images shown in the figures are provided in the Supplementary Information File.

\bibliography{sample}

\section*{Acknowledgements}
This work was supported by grants CPP2021-008562, CPP2023-01050, PID2023-149812OB-I00, PID2023-149812OB-I00, AGAUR (2022 DI 022 and GRC-2021 SGR 00113), UAB PPC2023\_575610, CNS2022-135512,  RYC2018-024346-I, CNS2022-135512, Torres Quevedo PTQ2023-013018 funded by MCIN$/$AEI$/$10.13039$/$501100011033 and by “European Union NextGenerationEU$/$PRTR”; and funded by Generalitat Valenciana CIAICO/2022/020 and by “ESF Investing in your future”.

\section*{Author contributions}
J.V: conceptualization, data curation, software development, writing, manuscript review and editing.
J.C: conceptualization, manuscript review and editing.
E.Z: software development, clinical data curation, manuscript revision.
I.L: numerical simulations generation, manuscript revision.
M.M: clinical data collection, manuscript revision. 
J.S: numerical simulations generation, manuscript revision.
J.R: clinical protocols, clinical data collection, manuscript revision.  
I.R: clinical protocols, clinical data collection, manuscript revision. 
L.M: clinical protocols, clinical data collection, manuscript revision. 
F.A: clinical protocols, clinical data collection, manuscript revision. 
A.M.C: conceptualization, manuscript review and editing.
M.S.G: manuscript revision.
I.H: conceptualization, software development, manuscript review and editing. 
All authors read and approved the final manuscript.

\section*{Competing interests}
Jorge Vicente-Puig is pursuing an industrial PhD co-supervised by Universitat Autònoma de Barcelona and Corify Care.
Ernesto Zacur and Jana Reventós are employees of Corify Care. 
Andreu M. Climent, Maria S. Guillem, and Felipe Atienza report board membership and ownership of equity or stocks in Corify Care. Ismael Hernández-Romero reports employment and ownership of equity or stocks in Corify Care. 
Lluis Mont reports honoraria as a consultant, lecturer, and Advisory Board from Boston-Scientific, Abbott, Johnson\&Johnson, and Medtronic, and is a shareholder of Galgo Medical SL.
Ivo Roca-Luque has received honoraria as a lecturer and consultant from Abbott and Biosense-Webster.
All other authors declare no known competing financial interests or personal relationships that could have appeared to influence the work reported in this paper.

\end{document}